\documentclass{svjour3} % onecolumn (standard format)
\smartqed
\usepackage{graphicx}
\usepackage{natbib}
\usepackage{amssymb}

\newcommand{\aap}{A\&A}
\newcommand{\apj}{ApJ}

\newcommand{\apjs}{ApJS}
\newcommand{\mnras}{MNRAS}
\newcommand{\pasj}{PASJ}
\newcommand{\nat}{Nature}
\newcommand{\pr}{Phys. Rep.}
\newcommand{\araa}{ARA\&A}
\newcommand{\mpla}{Mod. Phys. Lett. A}

\newcommand{\adspr}{Adv. Sp. Res.}
\newcommand{\qjras}{QJRAS}
\newcommand{\ssr}{SSR}

\newcommand{\lya}{Ly$\alpha$}

\newcommand{\ion}[2]{#1\,{\sc{#2}}}

\newcommand{\uno}{$^{-1}$}
\newcommand{\due}{$^{-2}$}

\newcommand{\xeus}{{\it XEUS}}
\newcommand{\chandra}{{\it Chandra}}
\newcommand{\xmm}{{\it XMM-Newton}}
\newcommand{\constellation}{{\it Constellation-X}}
\newcommand{\euve}{{\it EUVE}}
\newcommand{\rosat}{{\it ROSAT}}
\newcommand{\asca}{{\it ASCA}}
\newcommand{\dios}{{\it DIOS}}
\newcommand{\fuse}{{\it FUSE}}
\newcommand{\spear}{{\it SPEAR}}

\newcommand{\edge}{{\it EDGE}}
\newcommand{\hst}{{\it Hubble Space Telescope}}

\newcommand{\la}{\mathrel{\mathchoice{\vcenter{\offinterlineskip\halign{\hfil
$\displaystyle##$\hfil\cr<\cr\sim\cr}}}
{\vcenter{\offinterlineskip\halign{\hfil$\textstyle##$\hfil\cr<\cr\sim\cr}}}
{\vcenter{\offinterlineskip\halign{\hfil$\scriptstyle##$\hfil\cr<\cr\sim\cr}}}
{\vcenter{\offinterlineskip\halign{\hfil$\scriptscriptstyle##$\hfil\cr<\cr\sim\cr}}}}}
\newcommand{\ga}{\mathrel{\mathchoice{\vcenter{\offinterlineskip\halign{\hfil
$\displaystyle##$\hfil\cr>\cr\sim\cr}}}
{\vcenter{\offinterlineskip\halign{\hfil$\textstyle##$\hfil\cr>\cr\sim\cr}}}
{\vcenter{\offinterlineskip\halign{\hfil$\scriptstyle##$\hfil\cr>\cr\sim\cr}}}
{\vcenter{\offinterlineskip\halign{\hfil$\scriptscriptstyle##$\hfil\cr>\cr\sim\cr}}}}}

\journalname{SSRv}

\begin{document}

\title{Numerical simulations of the Warm-Hot Intergalactic Medium}
\titlerunning{WHIM simulations}

\author{Serena Bertone \and Joop Schaye \and Klaus Dolag}
\authorrunning{Bertone et al.}

\institute{
S. Bertone \at Astronomy Centre, University of Sussex, Falmer, Brighton BN1 9QH, United Kingdom \\
\emph{Present address:} Santa Cruz Institute for Particle Physics, University of California at Santa Cruz, 1156 High Street, Santa Cruz CA 95064, USA \\
\email{serena@scipp.ucsc.edu}
\and
J. Schaye \at Leiden Observatory, Leiden University, P.O. Box 9513, 2300 RA Leiden, The Netherlands \\
\email{schaye@strw.leidenuniv.nl}
\and
K. Dolag \at Max-Planck-Institut f\"ur Astrophysik, P.O. Box 1317, D-85741 Garching, Germany \\
\email{kdolag@mpa-garching.mpg.de}
}

\date{Received: 20 October 2007 ; Accepted: 22 November 2007}

\maketitle

\begin{abstract}

In this paper we review the current predictions of numerical
simulations for the origin and observability of the warm hot
intergalactic medium (WHIM), the diffuse gas that contains up to 50
per cent of the baryons at $z\sim 0$. During structure formation,
gravitational accretion shocks emerging from collapsing regions gradually heat
the intergalactic medium (IGM) to temperatures in the range $T\sim
10^5 - 10^7$ K. The WHIM is
predicted to radiate most of its energy in the ultraviolet (UV) and
X-ray bands and to contribute a significant fraction of the soft X-ray
background emission.  While \ion{O}{vi} and \ion{C}{iv} absorption systems arising
in the cooler fraction of the WHIM with $T\sim 10^5 - 10^{5.5}$ K are
seen in \fuse\ and \hst\ observations, models agree that current X-ray
telescopes such as \chandra\ and \xmm\ do not have enough sensitivity
to detect the hotter WHIM. However, future missions such as
\constellation\ and \xeus\ might be able to detect both emission lines
and absorption systems from highly ionised atoms such as \ion{O}{vii},
\ion{O}{viii} and \ion{Fe}{xvii}.

\end{abstract}

\keywords{large-scale structure of the universe, intergalactic medium, diffuse radiation, X-rays: diffuse background, quasars: absorption lines}

\section{Introduction}
\label{Introduction} 

The cosmic baryon abundance was first inferred by applying the theory
of primordial nucleosynthesis to observations of light element
abundances (e.g.\ \citep{olive2000}) and has been confirmed by observations 
of the cosmic microwave background radiation (CMB). A recent
measurement from the Wilkinson Microwave Anisotropy Probe reveals that
$\Omega_{\mathrm b} h^2 = 0.0223_{-0.0009}^{+0.0007}$, where the Hubble
parameter is $h=0.73_{-0.04}^{+0.03}$ \citep{spergel2007}. At $z> 2$ most of the baryons are thought to reside in the diffuse, photoionised intergalactic medium (IGM) with 
$T\sim 10^4 - 10^5$ K, traced by the \lya\ forest\footnote{Since the
diffuse intergalactic medium is highly ionised, the baryon
density inferred from \lya\ forest observations is inversely
proportional to the assumed ionisation rate. Since independent
constraints on the UV intensity are only accurate to a factor of few
at best, we cannot prove that most of the baryons reside in the
forest.} (e.g.\ \citealt{rauch1997}; \citealt{weinberg1997};
\citealt{schaye2001}). The gas 
traced by the \lya\ forest, together with the mass in galaxies, fully
accounts for the cosmic baryon abundance at $z\sim 2$ and confirms
the prediction of nucleosynthesis theories and the CMB
measurements.

However, when the mass of stars, interstellar gas and plasma in
clusters of galaxies at redshift $z\sim 0$ is considered, only a small
fraction of the mass budget can be accounted for
(e.g.\ \citealt{persic1992}; \citealt{fukugita1998};
\citealt{fukugita2004}). While the gas traced by \lya\ forest may
account for about a third of the low-$z$ baryons
(e.g.\ \citealt{danforth2007}), the majority of the baryons remain
invisible. Unless the baryon budget at high redshift has been
overestimated by two different measurements, a large amount of the low
redshift baryons must be ``missing''.

One of the basic predictions of cosmological, gas-dynamical simulations is the
distribution of baryons in the universe. How many baryons are locked in stars? How many reside in clusters and filaments? What is the state of the diffuse gas? Simulations of structure formation predict that the missing baryons problem is measure of our technological limitations rather than a real problem. In other
words, the baryons are out there, but we cannot see them because they reside in gas that is too dilute to be detected in emission and too hot to be traced by
\lya\ absorption.

The idea of gravitational heating of the intergalactic gas was first
suggested by \citet{sunyaev1972} and subsequently developed
(e.g.\ \citealt{Nath2001}; \citealt{furlanettoloeb2004}; 
\citealt{rasera2006}). While the largest structures in the universe form, 
the IGM is heated by gravitational shocks that
efficiently propagate from the collapsing regions to the surrounding
medium.  Simulations predict that gas compressed and heated by shocks
can reach temperatures of $10^8$ K in rich clusters of galaxies, while
filaments and mildly overdense regions are heated to temperatures in
the range $10^5$ to $10^7$ K (e.g.\ \citealt{Cen1999};
\citealt{dave2001}; \citealt{Cen2006}). The
latter are commonly known as the Warm-Hot Intergalactic Medium (WHIM),
which represents about half the total baryonic mass in the universe at
$z\sim 0$. 

At $T\sim 10^5 - 10^7$ K, the IGM is collisionally ionised and becomes
transparent to \lya\ radiation. As a consequence, the
\ion{H}{i} \lya\ forest does not trace the bulk of the gas mass at low
redshift. Although the shock-heated IGM emits radiation in the UV and in the
soft X-ray bands, the total energy radiated away by mildly overdense
gas is orders of magnitude too small to be detected by current
instruments. Similarly, the absorption from such a gas along the line
of sight to a bright X-ray source is too weak to be resolved by
current spectrographs (\citealt{richter2008} - Chapter 3, this volume).

In this Paper, we review the predictions of numerical simulations
for the properties and the observability of the WHIM.  In
Sect.~\ref{origin} we briefly mention the numerical techniques used to
simulate the IGM and we discuss the mechanisms that produce the WHIM,
namely heating by gravitational accretion shocks and other,
non-gravitational, heating processes. A more detailed description of
the numerical techniques themselves can be found in \citet{dolag2008}
- Chapter 12, this volume.  Sect.~\ref{background} describes the
contribution of the WHIM emission to the soft X-ray background. The
current predictions for the detectability of the WHIM in emission and
absorption are discussed in Sect.~\ref{emission} and in
Sect.~\ref{absorption} respectively. Sect.~\ref{sz} briefly reports on
the effect of the WHIM on the CMB radiation and Sect.~\ref{bfield}
discusses the magnetisation of the IGM. We draw our conclusions in
Sect.~\ref{conclusion}.

\section{Origin and properties of the WHIM}
\label{origin}

\subsection{Numerical techniques}
\label{simulations}

In this Subsection, we briefly discuss the numerical techniques that
have been used to investigate the IGM and the WHIM in
particular. However, we refer the reader to the review of
\citet{dolag2008} - Chapter 12, this volume for a thorough description
of specific simulation techniques.

The most common approach to model the WHIM is through
cosmological hydrodynamical simulations. The strength of
hydrodynamical simulations is that the basic gas dynamical processes
that determine the evolution of the diffuse gas are directly
simulated and that it is possible to make virtual observations of the
WHIM with existing or future instruments. However, \emph{ad hoc}
empirical prescriptions are needed 
to include processes such as galactic winds and feedback from active
galactic nuclei 
(AGN), whose physics is still poorly understood and in any
case would require a resolution far beyond the current standards
to be included self-consistently.  Hydrodynamical simulations of the
WHIM have typically used either Smoothed Particle Hydrodynamics  (SPH, e.g.\ \citealt{hellsten1998}; \citealt{croft2001}; \citealt{dave2001}; \citealt{Yoshikawa2004}; \citealt{Yoshikawa2006}; \citealt{dolag2006}), a uniform grid (e.g.\ \citealt{Cen1999}; \citealt{Cen2006}), or
Adaptive Mesh Refinement (AMR, e.g.\ \citealt{kravtsov2002}) to solve
the gas dynamics.  

Several properties of the WHIM, including the impact of gravitational
and non-gravitational heating processes on the IGM, and the
observability of emission lines and absorption systems in UV and X-ray
spectra, have been successfully investigated with analytical and
semi-analytical techniques (\citealt{perna1998}; \citealt{pen1999};
\citealt{Nath2001}; \citealt{Valageas2002}; \citealt{viel2003};
\citealt{furlanetto2004}; \citealt{viel2005};
\citealt{furlanetto2005}; \citealt{rasera2006}).

\subsection{Gravitational heating}

\begin{figure}
\centering \includegraphics[width=0.9\textwidth]{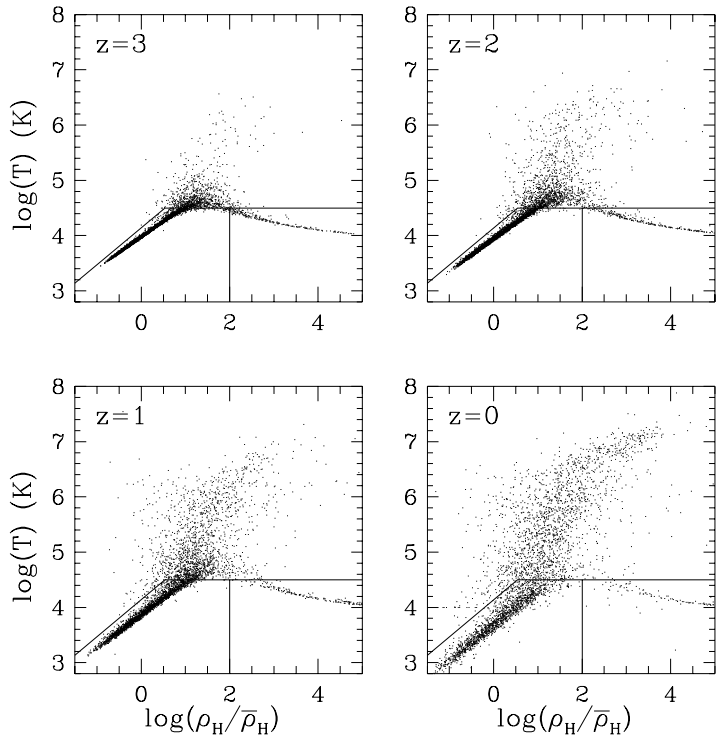}
\caption{ The temperature vs. overdensity plane at $z=$3, 2, 1, and
  0. The gas mostly resides in three phases: a cool, photoionised,
  diffuse IGM (tail to lower left); the warm-hot shocked IGM; and gas
  condensed into stars and star forming clouds (tail to lower
  right). The solid lines indicate an approximate separation between
  these phases. From \citet{dave1999}.}
\label{dave}
\end{figure}

Theoretical predictions and numerical simulations agree that the
heating of the IGM is fuelled by the gravitational accretion shocks emerging
during the collapse of cosmic structures (e.g. \citealt{Cen1999};
\citealt{dave2001}; \citealt{Nath2001}; \citealt{Valageas2002};
\citealt{kang2005}; \citealt{Cen2006}; \citealt{bykov2008} - Chapter 7, this volume).  
Simulations suggest that the distribution of the warm-hot gas
closely reflects the distribution of the shock waves,
and therefore the underlying density field.  The role of shock waves
in heating the IGM has been investigated using cosmological
hydrodynamical simulations by \citet{kang2005}, who find that the
fastest shocks can be found around clusters and groups of galaxies,
while the slowest shocks usually propagate across low density regions
like sheets and filaments. This reflects the fact that the infall
velocity is a measure of the depth of the potential
well. \citet{furlanetto2005} reach similar conclusions using
analytical techniques and find that the shocks with the highest
temperatures are associated with the most massive virialised objects,
up to distances larger than their virial radii.

The WHIM gas is organised in a complex structure of filaments and
sheets, that generally reflects the dynamics of structure
formation. According to \citet{dolag2006}, the gas density in
filaments is typically $10-100$ times the mean density and varies more
smoothly than in clusters. Filaments have a coherence length of about
5 Mpc, although some can be as long as 25 Mpc, and a diameter of about $3-5$ Mpc. The detailed thermal
structure of the WHIM strongly depends on the direction of propagation
of the accretion shocks, while the velocity field appears to be
orthogonal to the filaments at large distances from galaxy clusters
and aligned to the filaments at small distances.

Most simulations predict that between 30 and 50 per cent of all the
baryons in the low redshift universe have temperatures in the interval
$10^5 - 10^7$ K (e.g. \citealt{hellsten1998}; \citealt{Cen1999};
\citealt{dave1999}; \citealt{dave2001}; \citealt{Cen2006}).
Fig. \ref{dave} shows the distribution of the gas in the
temperature-density plane at $z=$3, 2, 1, and 0. Hot gas with $T>10^7$
K resides in the hottest cores of groups and in large clusters of
galaxies. Most of the gas, however, is at temperatures $T<10^7$ K and
at $z=0$ is about evenly spread between the WHIM and the cooler,
mostly photo-ionised IGM. At higher redshifts, a larger fraction of
the baryons resides in the photo-ionised IGM.  \citet{kang2005} find
that, besides the gas shock-heated to $T=10^5 - 10^7$ K, a significant
amount of mass, distributed mostly along sheet-like structures, is
shock-heated to $T <10^5$ K by shock waves propagating with velocities
$v_{\textrm{sh}} < 150$ km s\uno. This mostly photo-ionised gas is
responsible for the majority of the \ion{O}{v} and \ion{O}{vi} absorption lines,
but contributes little to the total \ion{O}{v} and \ion{O}{vi} emission, which is
instead produced by warmer, collisionally ionised ions.

\subsection{Non-equilibrium calculations}

Simulations that include non-equilibrium processes have been performed
by \citet{teyssier1998}, \citet{CenFang2006} and
\citet{Yoshikawa2006}.  In most hydrodynamical simulations, it is
routinely assumed that diffuse gas is in ionisation equilibrium
(\citealt{Cen1999}; \citealt{dave2001}; \citealt{Yoshikawa2003};
\citealt{Yoshikawa2004}; \citealt{kang2005};
\citealt{roncarelli2006}). In general, ionisation equilibrium is
nearly achieved in the centres of galaxy clusters and in photo-ionised
regions
with $T< 10^5$ K. However, this may not happen in the outer
regions of clusters, in groups, and in the WHIM \citep{Yoshikawa2003}.

Similar conclusions are reached by \citet{CenFang2006} and
\citet{Yoshikawa2006}. \citet{CenFang2006} claim that observational
results for the abundance of \ion{O}{vi} absorption lines are better
reproduced by models including non-equilibrium processes.
\citet{Yoshikawa2006} find that a significant fraction of the WHIM at
$z=0$ is not in ionisation equilibrium, as do
\citet{Yoshida2005}. However, \citet{Yoshikawa2006} calculate that
although this strongly affects the relative abundance of ions such as
\ion{O}{vi}, \ion{O}{vii} and \ion{O}{viii}, it does not have a significant effect on the
observable signatures of the WHIM in emission or absorption.

When gas is infalling onto a collapsing structure and is heated by
gravitational accretion shocks, most of its energy is carried by ions, while
electrons can only gain thermal energy through collisions with
ions. The temperatures of ions and electrons can therefore be very
different and converge only if energy can be efficiently transferred
from ions to electrons. If only Coulomb interactions are considered,
then the relaxation timescale can be 
shorter than the age of the universe for the WHIM (\citealt{Yoshikawa2003};
\citealt{Yoshida2005}). Since
the relative abundance of metal ions and the plasma emissivity are
sensitive to the electron temperature, this would have consequences
for the observability of the WHIM. 

\begin{figure}
\centering \includegraphics[width=10cm]{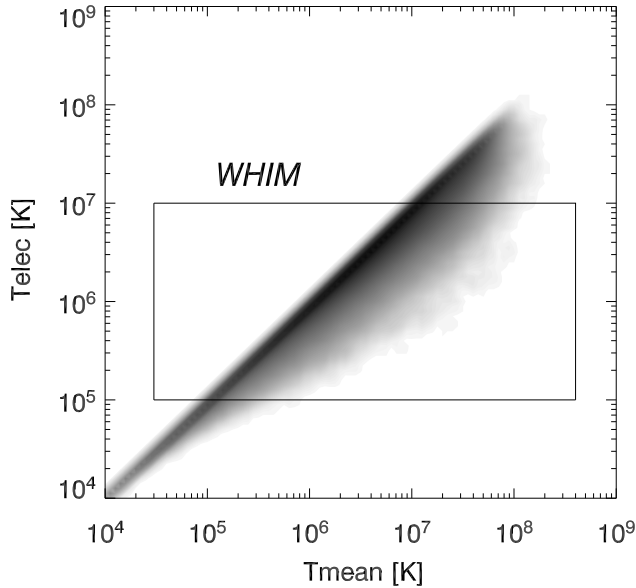}
\caption{Electron temperature versus the mean gas temperature. The
  grey scale shows the number density of gas particles and the
  rectangular box encloses the WHIM gas, with
  $10^5<T_{\textrm{elec}}<10^7$ K. From \citet{Yoshida2005}.}
\label{yoshida2005}
\end{figure}

\citet{Yoshida2005} introduce a two-temperature model for the thermal
evolution of an ionised gas. Their model includes an explicit
treatment for collisional relaxation processes between electrons and ions and
allows one to follow the temperature of free electrons separately from the
temperature of ions. They find that at $z=0$ most WHIM
gas shows a two-temperature structure. Fig. \ref{yoshida2005} shows
the electron temperature in the two-temperature model compared to the
mean temperature of the gas at $z=0$ for the simulations of
\citet{Yoshida2005}. Clearly, the electron temperature of the WHIM is
significantly lower than the mean gas temperature, when
non-equilibrium processes are taken into account. \citet{Yoshida2005}
calculate that the relative abundance of \ion{O}{vi}, \ion{O}{vii} and
\ion{O}{viii} predicted by a two-temperature model can be about an order of
magnitude different from that predicted by a one-temperature
model. This yields similarly different values for the line emission of
ions.

However, from supernova remnants we know that
plasma waves are typically much more efficient than collisional
processes in transferring energy from ions to electrons
(e.g.\ \citealt{Rakowski2005}) and these effects have not yet been
included in simulations of the WHIM. \citet{Yoshida2005} may therefore
have overestimated the importance of the two-temperature structure of
the post-shock gas.

\subsection{Non-gravitational physics}

Several studies have investigated whether sources of non-gravitational
heating might significantly affect the thermal state of the IGM. Simulations by \citealt{Cen2006}
predict that non-gravitational processes, such as galactic winds
and X-ray emission from galaxies and quasars, contribute no more than
20 per cent of the energy required to heat the IGM, in agreement with
previous results from \citet{dave2001} and \citet{Nath2001}.

A powerful test to constrain the contribution of non-gravitational
processes is the comparison of the observed intensity of the soft
X-ray background to the predictions of numerical simulations. Since most of the observed background emission has been resolved into
point sources (\citealt{hasinger1993}; \citealt{worsley2005}), the
comparison of the predictions of hydrodynamical simulations with
observations can put tight constraints on the amount of the observed
background emission that can be attributed to the diffuse
intergalactic gas.

In general, the intensity of the soft X-ray background predicted by
simulations that include only gravitational heating processes strongly
exceeds the observed emission from high density, collapsed regions
(e.g. \citealt{Nath2001}; \citealt{Bryan2001}; \citealt{dave2001};
\citealt{Cen2006}). The emission is primarily due to Bremsstrahlung
and reflects the cooling of gas in groups and clusters of galaxies.
Analytical calculations by \citet{pen1999} and \citet{wu2001} find an order of magnitude more background light than predicted by simulations, owing to the importance of gravitational pre-heating.
They suggest that additional heating from non-gravitational processes might be necessary to ``puff up'' and hence decrease the emission from collapsed regions and reconcile the theoretical predictions with the observations. However, they note that non-gravitational heating does not seem to be relevant for uncollapsed regions like filaments and voids.

\citet{voit2001} argue that non-gravitational processes can help to establish an entropy floor in high density regions. The lowest entropy material will drop out and form stars unless it is heated through feedback from star formation. They suggest that an entropy floor of $\sim 100$ keV cm\due\ is 
required to reconcile the theoretical results with observations.
\citet{Bryan2001} demonstrate further that radiative cooling and
additional heating from feedback can efficiently suppress the X-ray
background, while substantially modifying the characteristic spectral
shape of the background itself.

\section{WHIM contribution to the soft X-ray background}
\label{background}

The extended spatial distribution over cosmological scales makes the
WHIM an essential contributor to the diffuse X-ray background
radiation. The most intense X-ray emission comes from relatively
compact sources such as clusters, active galactic nuclei and
galaxies. The gas in these objects has high emissivity power, but
accounts for only a small fraction of the baryon mass in the
universe. The emissivity of the WHIM is usually a few orders of
magnitude lower, because of its lower density and
temperature. However, the WHIM accounts for up to half the baryonic
mass at $z\sim 0$ and its integrated emission over cosmological scales
might be roughly comparable to that of dense regions and compact
objects.

\begin{figure}
\hbox{
\includegraphics[width=0.47\textwidth]{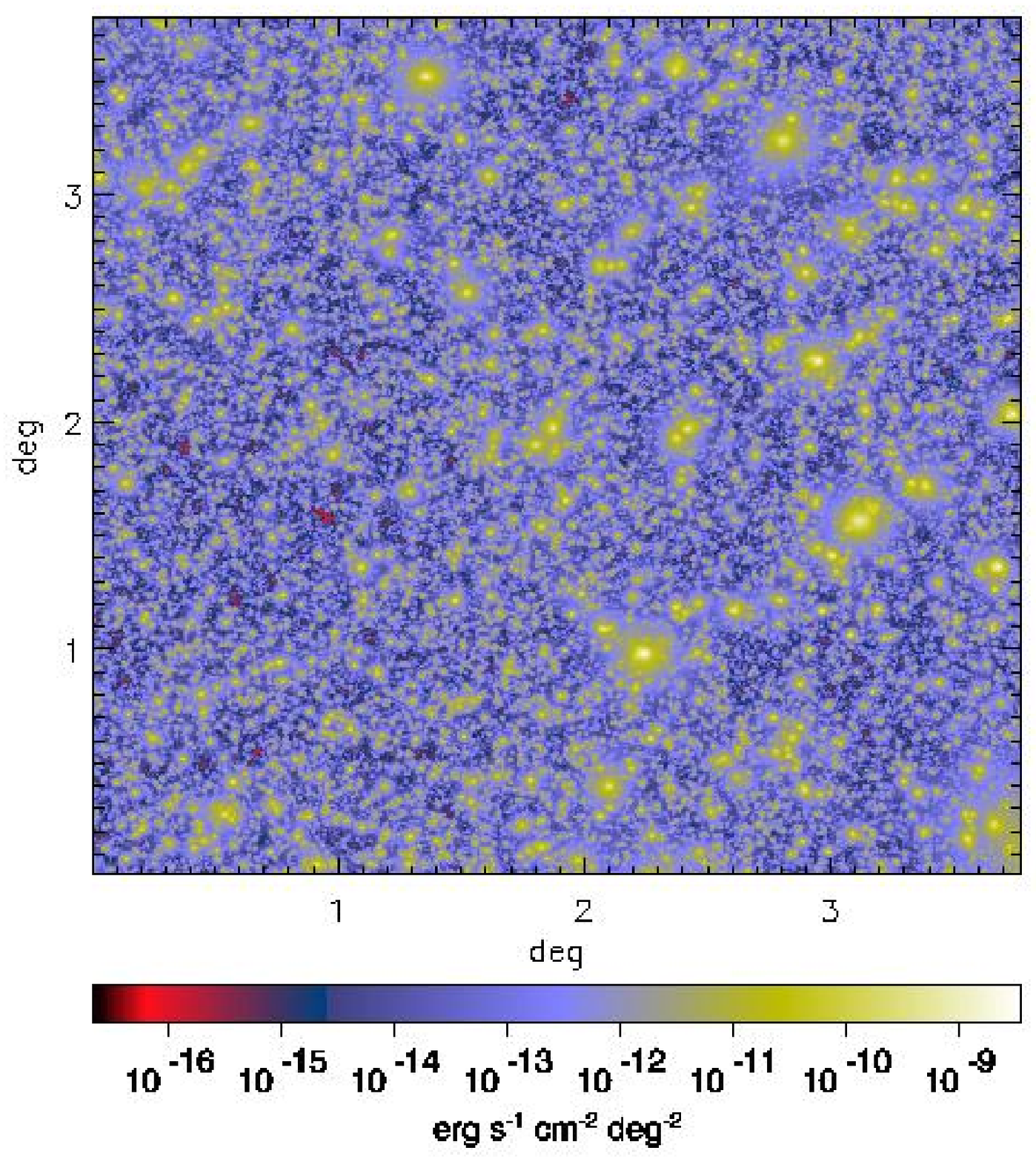}
\hspace{4mm}
\includegraphics[width=0.47\textwidth]{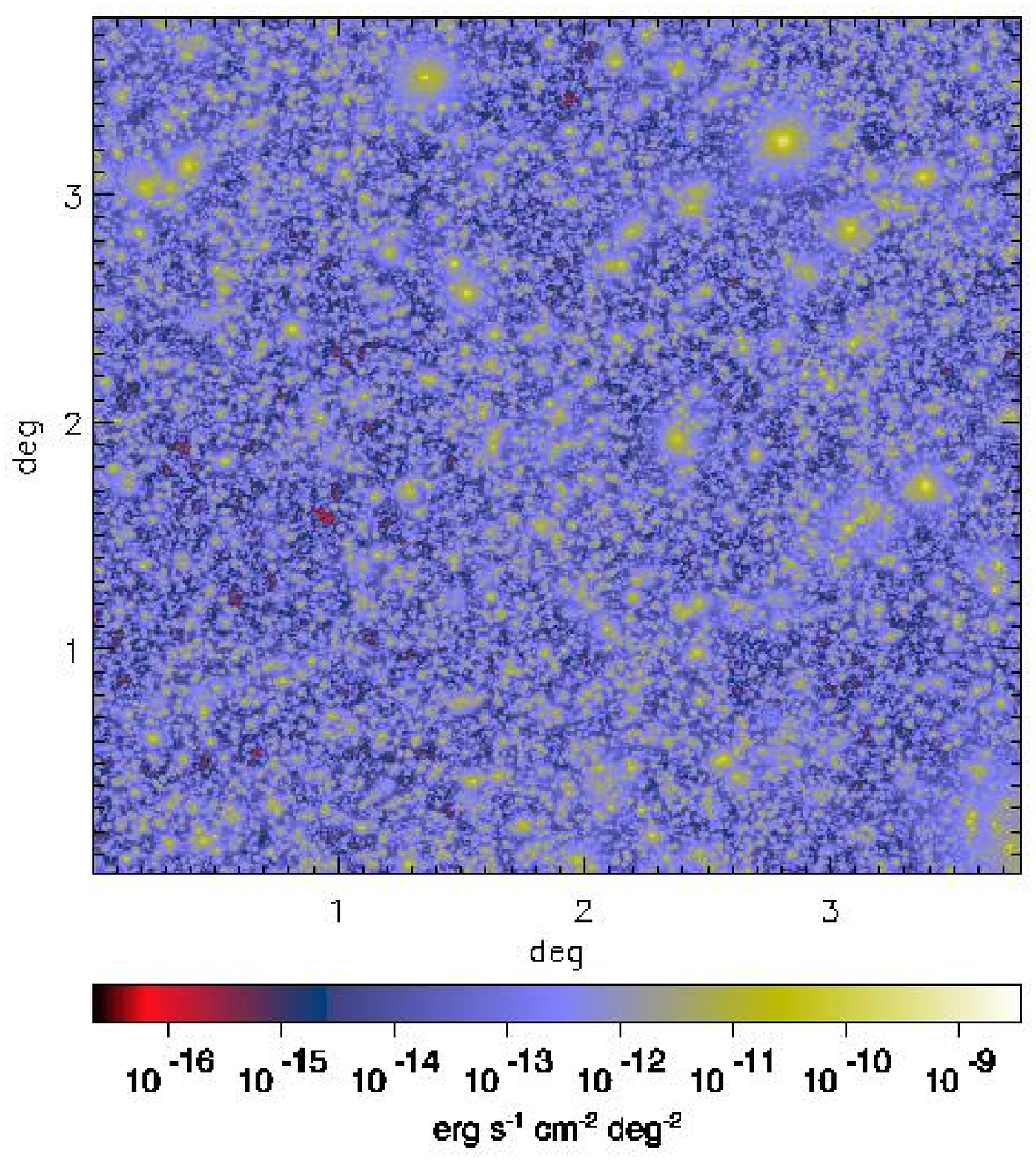}}
\hbox{
\includegraphics[width=0.47\textwidth]{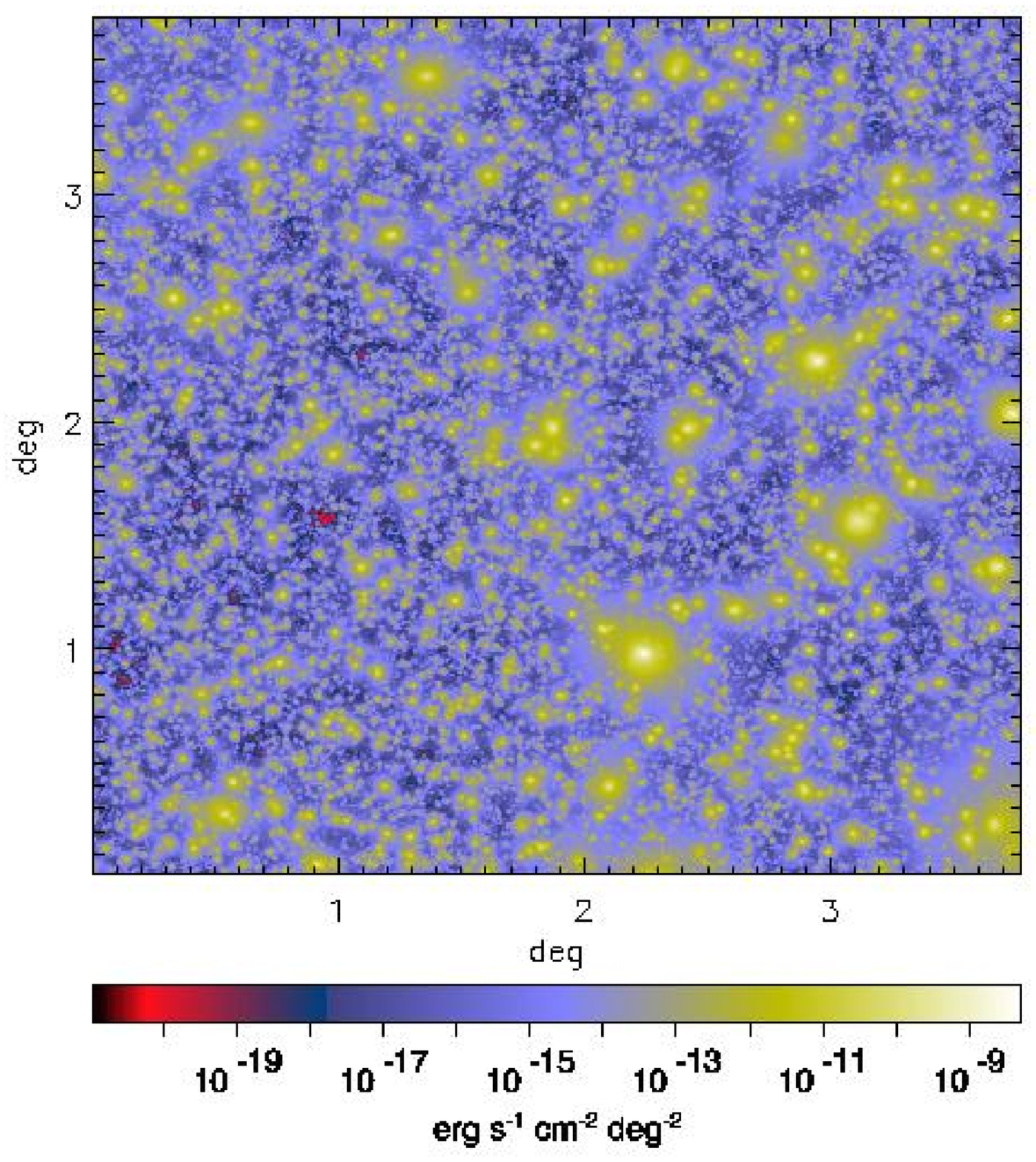}
\hspace{4mm}
\includegraphics[width=0.47\textwidth]{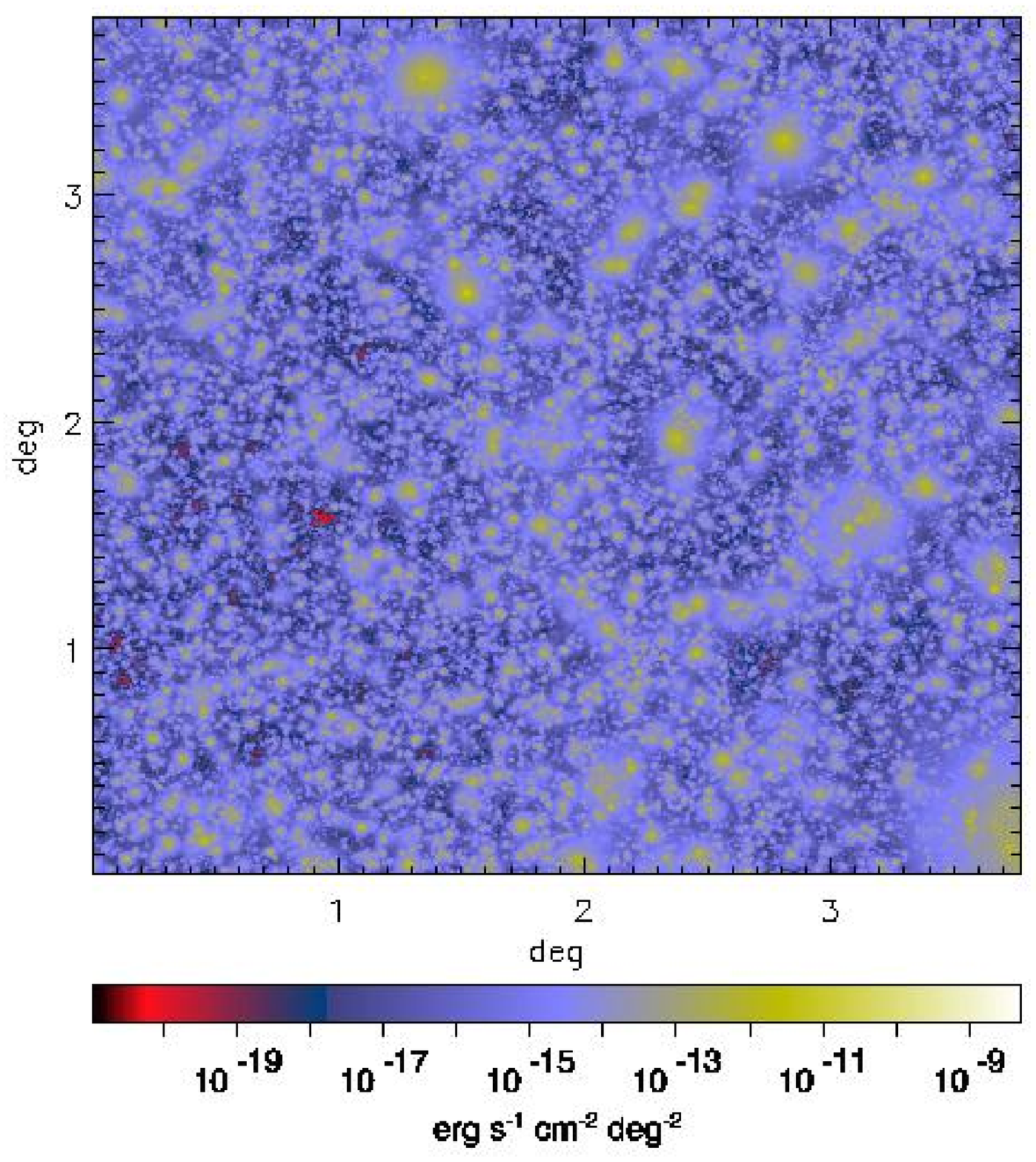}}
\caption{Maps of the soft ($0.5-2$ keV, upper panels) and hard ($2-10$
  keV, lower panels) X-ray intensity obtained by considering all the
  gas particles (left panels), and only the gas particles with
  temperatures in the range $10^{5}<T<10^{7}$ K (WHIM, right
  panels). The maps are 3.78 deg on a side, with a pixel size
  $(1.66~\textrm{arcsec})^2$. The two maps show results for the same
  realisation of the past light-cone. From \citet{roncarelli2006}.}
\label{roncarelli}
\end{figure}

Analytical calculations \citep{Valageas2002} and numerical simulations
predict that the WHIM contributes between a few per cent
\citep{kang2005} and 40 per cent (\citealt{croft2001}; \citealt{roncarelli2006}) of the total X-ray background emission, consistent with the observational upper
limit of $(1.2 \pm 0.3) \times 10^{-12}$ erg\,s\uno\,cm\due\,deg\due\ \citep{worsley2005}.  Because of the softness of the spectrum, the WHIM produces only between 4 per cent \citep{croft2001} and 10 per cent \citep{roncarelli2006} of the total hard X-ray emission. According to the simulations of \citet{roncarelli2006}, about 90 per cent of the X-ray background emission is produced at $z<0.9$.
Fig. \ref{roncarelli} shows simulated maps of the soft ($0.5-2$ keV,
upper panels) and hard ($2-10$ keV, lower panels) X-ray emission from a
region 3.78 degree on a side. The left panels show the total emission
from the gas, while the right panels show the contribution of the
WHIM.

\begin{figure}
\centering \includegraphics[width=0.9\textwidth]{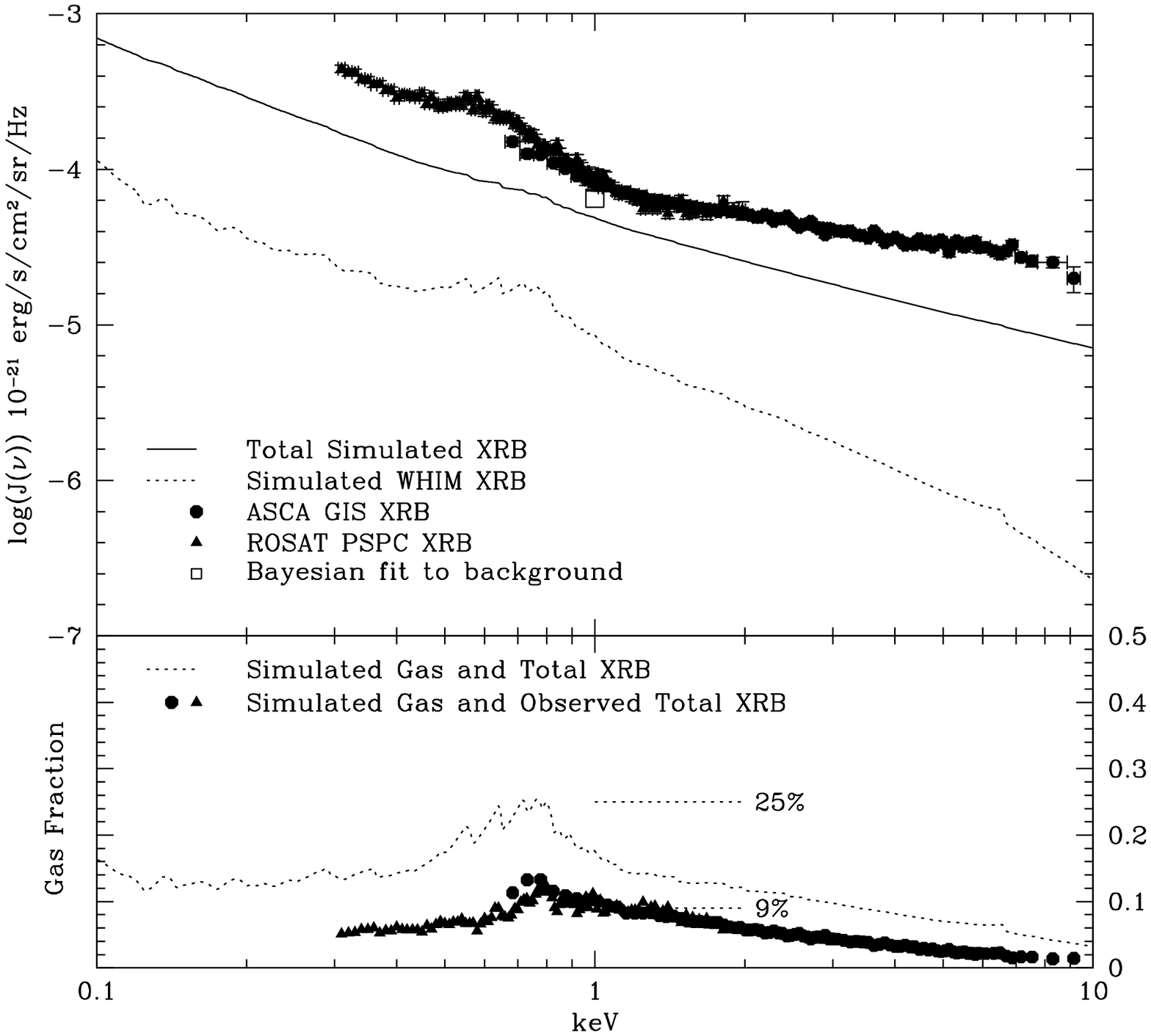}
\caption{The predicted integrated X-ray background (solid curve) and the
  predicted contributions from the IGM (dashed curve) and WHIM alone (dotted
  curve).  The 
  observations of the XRB spectrum from the ASCA LSS region
  \citep{Miyaji2000} and ROSAT PSPC fields are shown along with the
  Bayesian fit to observations of the X-ray background
  \citep{barcons2000}. The fractional contribution of the WHIM to the
  total simulated and observed backgrounds is plotted in the lower
  panel. \citet{Phillips2001} argue that the factor of two difference 
  between the normalisation of the simulated background and that of the 
  observed background may be due to the low value of $\Omega_{\mathrm b}$ 
  assumed in the simulations or to the underestimate of the observed AGN 
  formation rate used to compute the spectrum. From \citet{Phillips2001}.}
\label{phillips}
\end{figure}

\citet{Phillips2001} estimate that in hydrodynamical simulations the
WHIM gas at $z=0$ is responsible for about $5-15$ per cent of the
total extragalactic surface brightness in the energy range $0.5-2$ keV
(Fig. \ref{phillips}), corresponding to a flux of $0.24 \times
10^{-12}$ erg\,s\uno\,cm\due\,deg\due. The WHIM line emission peaks in
the $0.5-0.8$ keV energy range, where observations by \asca\ and
\rosat\ show a spectral bump, as shown in
Fig. \ref{phillips}. \citet{Phillips2001} argue that the bump is due
to WHIM emission, and in particular to metal line emission.
Similarly, \citet{kravtsov2002} produce simulations that faithfully
reconstruct the local universe and find that about $5-10$ per cent of
the X-ray background at energies of about 1 keV may be contributed by
the WHIM.  \citet{Ursino2006} find a contribution of up to 20 per cent
in the energy range $0.37 - 0.925$ keV, most of which is emitted by
filaments at redshifts between 0.1 and 0.6. Dense and bright filaments
are present in about 10 per cent of a generic field of view, and can
account for more than 20 per cent of the total emission. In such
cases, redshifted emission lines from highly ionised atoms should be
clearly detectable.

\subsection{Soft X-ray excess radiation in galaxy clusters}

Soft X-ray excess radiation in clusters has been detected in
observations by the Extreme Ultraviolet Explorer (\euve), \rosat\ and
\xmm\ (\citealt{lieu1996}; \citealt{bonamente2001};
\citealt{Finoguenov2003}; see \citealt{durret2008} - Chapter 4, this volume for an overview).  The temperature of the ICM is usually estimated by
assuming a one-temperature plasma model for the diffuse gas. The
soft X-ray flux  calculated from the observed X-ray emission of clusters,
however, often appears to be above the expected thermal
contribution. It has been suggested that the excess radiation may be
due to foreground emission from the WHIM, or from warm diffuse gas in
the clusters.

\citet{cheng2005} investigate the origin of this excess with a set of
cluster simulations. They find that the presence of WHIM is not
necessary to explain the observations, because the excess emission is
produced by extremely dense gas associated with merging substructures
within the cluster. In clusters where this gas is present, the
temperature of the inner regions is usually $30-50$ per cent above
that predicted by the one-temperature plasma model.
\citet{dolag2006} find that, since the WHIM surface brightness in
X-rays reaches at most $10^{-16}$ erg\,s\uno\,cm\due\,arcmin\uno,
projection effects of WHIM filaments in front of galaxy clusters can
account for no more than 10 per cent of the cluster emission. This may
partly explain the soft X-ray excess, but clearly does not fully
account for it. Similar conclusions have been reached by
\citet{mittaz2004}.

\section{Line emission}
\label{emission}

Most of the energy radiated by the WHIM in the UV and soft X-ray bands
is emitted through emission lines.  The detection of emission lines
from the WHIM is key to understanding the properties of the gas
itself. The relative intensities of emission lines from highly ionised
metals in the WHIM gives 
information about the temperature of the emitting gas, while the
absolute intensities reflect the density and the degree of metal
enrichment of the gas. 

The relative abundances of ionised atoms is determined by the
temperature and, if photo-ionisation plays a role, the density of the
WHIM. Ions whose abundance peaks at the
highest temperatures, such as \ion{Fe}{xxv} and \ion{Fe}{xxvi}, are concentrated in
the hottest, highest density regions, while ions that peak at lower
temperatures, such as \ion{O}{vii}, \ion{O}{viii} and \ion{Fe}{xvii}, are more widespread
and originate in groups and filaments \citep{Fang2002}.  The detection
of WHIM emission from the intracluster medium of filaments, groups and
cluster with different masses helps to understand the impact of
feedback in these structures \citep{pierre2000}.

\subsection{Soft X-rays}

\begin{figure}
\centering \includegraphics[width=0.9\textwidth]{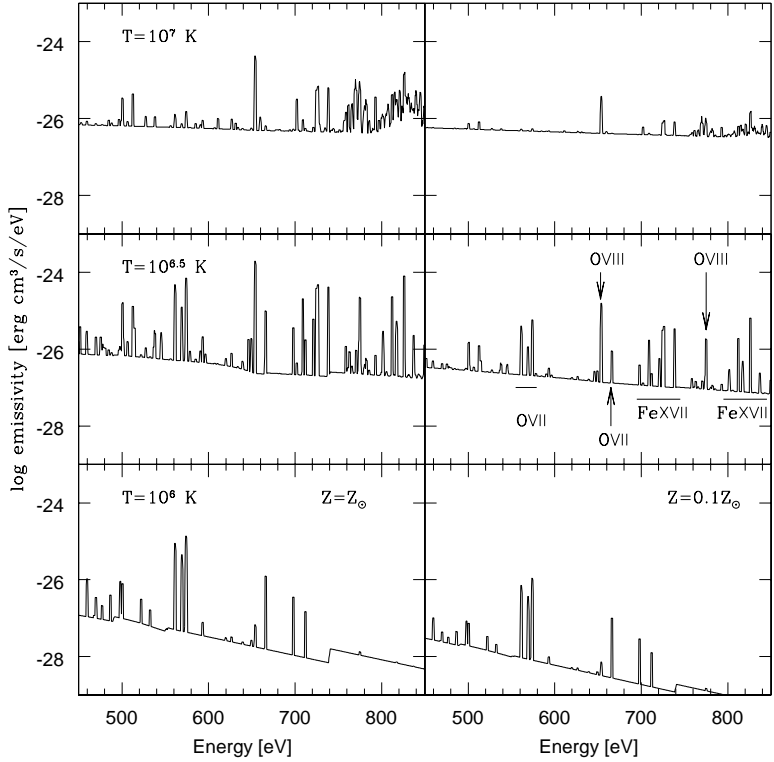}
\caption{Template spectra of a collisionally ionised plasma with
  temperature $T=10^6$ K (lower panels), $10^{6.5}$ K (middle panels),
  and $10^7$ K (upper panels). Spectra for metallicity $Z=Z_{\odot}$
  and $Z=0.1Z_{\odot}$ are shown in the left and right panels,
  respectively.  From \citet{Yoshikawa2003}.}
\label{yoshikawa2003}
\end{figure}

\citet{Yoshikawa2003} and \citet{Yoshikawa2004} find that the
temperature of the IGM can be estimated if both the \ion{O}{vii} and
\ion{O}{viii} ions are simultaneously observed.  However, if only one line
is detectable, then the line cannot be identified
unambiguously. Fig. \ref{yoshikawa2003} shows 
template spectra of a collisionally ionised plasma with temperature
$T=10^6$ K, $10^{6.5}$ K, and $10^{7}$ K, and with metallicity
$Z=Z_{\odot}$ and $Z=0.1Z_{\odot}$, which display the contribution of
oxygen ions as a function of temperature. Clearly, the detectability
of the line emission depends strongly on the metallicity. 

It is still controversial if the WHIM can actually be detected in
emission by current and future X-ray telescopes.  \citet{voit}
calculate that the virialised regions of groups and clusters cover
about a third of the sky. This is a potential problem in the
identification of X-ray emission from the WHIM in filaments and
distant groups, because it creates source-confusion and strongly
contaminates the observations.  \citet{pierre2000} predict that
\xmm\ should be able to detect diffuse emission from strong filaments
at $z \sim 0.1$, although no such detections have been confirmed so
far.  As \citet{Fang2005} calculate, X-ray emission lines from the
WHIM are hard to detect in high density regions and in voids. The
emission from clusters and groups is dominated by the hot intracluster
medium, while in voids the X-ray spectrum is completely dominated by
radiation from background AGN and by the Galactic foreground. In both
cases, the emission of the WHIM would be too low to be distinguished
from the dominant component. On the other hand, it might be possible
to resolve at least a few strong emission lines from the WHIM in
filamentary regions, despite the fact that most of the emission is
contributed by AGN and by the Galactic foreground.

\citet{Fang2005} argue that planned telescopes such as
\constellation\ and the X-ray Evolving Universe Spectroscopy mission
(\xeus) 
have little chance to see a significant signal for the WHIM emission,
because their effective collecting area is too small to detect diffuse
line emission from an extended area. Ideally, a mission to detect line
emission from the WHIM should have a large field of view, coupled with
a high resolving power and spectral resolution.  \citet{Yoshikawa2003}
and \citet{Yoshikawa2004} use large scale simulations and simulations
that reproduce the local universe to produce mock X-ray spectra of the
WHIM emission. The simulated maps of the local universe
\citep{Yoshikawa2004} demonstrate that WHIM gas in filamentary
structures can be successfully imaged in targeted observations by the
proposed Diffuse Intergalactic Oxygen Surveyor (\dios), an X-ray
telescope of new generation which combines a large field of view with
high spatial and spectral resolution. \dios\ and several other
proposed missions are discussed in more detail in \citet{paerels2008} - Chapter 19, this volume.

\subsection{UV}

While \ion{O}{vii} and \ion{O}{viii} lines are emitted by gas with $T\ga 10^6$
K, line emission from the \ion{O}{vi} $\lambda \lambda 1032, 1038$~\AA\ and
\ion{C}{iv} $\lambda \lambda 1548, 1551$~\AA\ doublets traces the coolest
fraction of the WHIM, with $T \sim 10^5 - 10^{5.5}$ K.  For the bulk
of the WHIM mass, the line 
strengths are predicted to be a few orders of magnitude lower than the
background \citep{furlanetto2004}. \citet{kravtsov2002} estimate the
intensity of the \ion{O}{vi} emission line at $\delta \sim 10$ and column
densities $N_{\mbox{O\,\scriptsize{VI}}} \sim 10^{14}$ cm\due\ to be $I\sim 1-10$ photons
cm\due\ s\uno\ sr\uno, which is an order of magnitude below the
sensitivity of past and current telescopes such as the Far Ultraviolet
Spectroscopic Explorer (\fuse) and \spear.

However, \citet{furlanetto2004} predict that \ion{O}{vi} and \ion{C}{iv} line
emission could be detected above the background from enriched dense
regions with sizes of about $50-100$ kpc. \citet{furlanetto2004} argue
that the connection  
between the line emission and the enrichment by galactic winds may
help understand the chemical enrichment history of the IGM and the
physics of winds.

\section{Absorption}
\label{absorption}

The X-ray forest is the high-energy counterpart of the
\lya\ forest. While the \lya\ forest is produced by cool gas absorbing
\lya\ photons in the optical spectra of high redshift quasars, the
X-ray forest is the result of absorption by highly ionised atoms in
the spectra of bright X-ray sources (\citealt{richter2008} - Chapter 3, this volume).

The detection of the X-ray forest in high resolution spectra of
quasars and gamma ray bursts (GRB) is possibly one of the most
powerful tools to investigate the properties of the WHIM in the low
redshift universe.  For example, the number density of absorbers gives
an estimate of the baryon density times the metallicity of the IGM,
while the ratio between the \ion{O}{vii} and the \ion{O}{viii} line strengths
gives information on the distribution of the gas temperature and
density.

A thorough knowledge of the physical state of the WHIM, as
\citet{furlanetto2005} point out, helps to understand the mechanisms
that have determined its evolution in a cosmological context.
\citet{furlanetto2005} use analytic techniques to investigate how the
accretion shocks that heat up the IGM during the formation of cosmic
structures can give rise to absorption systems such as \ion{O}{vi} and
\ion{O}{vii}.  They find that the observed column densities of absorbers are
predicted by post-shock cooling models, when both slow and fast
cooling channels are taken into account. The observed number density
of absorbers can be realistically reproduced if each virialised
structure is surrounded by a network of shocks with a total cross
section a few times the size of the virialised region.  In this
scenario, fast cooling would produce stronger \ion{O}{vii} absorbers
associated to \ion{O}{vi} systems than expected from models of collisional
ionisation equilibrium. \ion{O}{vii} absorption systems should be the most
common because the abundance of the \ion{O}{vii} ion dominates over a large
temperature range.

\subsection{Soft X-rays}

The observability of the WHIM in the X-ray forest by current and
future facilities has been a very popular topic in recent years
(\citealt{richter2008} - Chapter 3, this volume). A variety of
predictions has been provided by hydrodynamical simulations
(e.g. \citealt{hellsten1998}; \citealt{kravtsov2002};
\citealt{Chen2003}; \citealt{CenFang2006}; \citealt{Kawahara2006}) and
semi-analytic models (e.g. \citealt{perna1998}; \citealt{viel2003};
\citealt{viel2005}).

\citet{kravtsov2002} find that the regions of the highest column
densities of \ion{O}{vii} and \ion{O}{viii} correspond to the high density regions
in and around galaxy groups. The gas in these regions, with $T=10^6$ K
and $\delta \sim 100$, should produce the strongest \ion{O}{vii} and
\ion{O}{viii} absorption lines, with equivalent width $W> 100$ km
s\uno\ \citep{hellsten1998}. These lines should occur on average once
per sight line in the redshift range $z=0-0.3$ (\citealt{hellsten1998};
\citealt{CenFang2006}).  \citet{CenFang2006} predict  abundances
for \ion{O}{vi}, \ion{O}{vii}, and \ion{O}{viii} lines in the range $50-100$ per unit
redshift at $W = 1$ km s\uno, decreasing to $10-20$ per unit redshift at
$W = 10$ km s\uno.

\citet{Chen2003} suggest that the most promising strategy to find
\ion{O}{vii} and \ion{O}{viii} absorbers in the X-ray forest would be to search at
the redshifts of known \ion{O}{vi} absorbers. However, since
\ion{O}{vi} absorption traces gas at somewhat lower temperatures than
\ion{O}{vii} and \ion{O}{viii}, only future missions like \constellation\ and
\xeus\ will be able to find absorption systems at higher temperatures.

Most predictions agree that \chandra\ and \xmm\ do not have enough
sensitivity to detect the WHIM in absorption.  \citet{Chen2003} claim
that a few strong \ion{O}{vii} and \ion{O}{viii} absorption systems might be
within reach of \chandra\ and \xmm, but the probability to detect such
lines in random lines of sight is less than 5 per cent
\citep{kravtsov2002}.

Future telescopes such as \constellation\ and \xeus\ might have a
better chance to detect the WHIM in absorption, thanks to higher
spectral resolution and larger collecting areas.
\constellation\ should have a 50 per cent probability to detect the
strongest \ion{O}{vii} and \ion{O}{viii} absorption lines
\citep{kravtsov2002}. Because of its high sensitivity at high
energies, it should also be able to detect absorption from ions such
as \ion{Ne}{ix} and \ion{Fe}{xvii}, whose intensity peaks at about the same
temperature as \ion{O}{vii}, and \ion{Fe}{xx} lines, which appear for $T\ga
10^7$ K \citep{hellsten1998}.  \citet{perna1998} predict that one day
of integration time will allow \constellation\ and \xeus\ to detect a
few \ion{O}{viii} absorption lines per unit redshift in the spectra of X-ray
bright quasars.  According to \citet{viel2003}, \constellation\ will
detect about 6 absorption lines per unit redshift with $W>10$ km
s\uno, while \xeus\ could detect up to 30 lines with $W>1$ km
s\uno\ per unit redshift.  Under the assumption of constant IGM
metallicity, \citet{Kawahara2006} estimate that \xeus\ will be able to
detect on average two \ion{O}{vii} absorption systems with $\sigma > 3$
along any line of sight to bright quasars at $z\la 0.3$ in an 8
hours exposure. 

It should be noted, however, that the characteristics
of the proposed facilities have been evolving and that some of the
above predictions are therefore likely too optimistic.

\subsection{UV}

\citet{Cen2001} find that about 20--30 per cent of the WHIM at $z=0$,
or equivalently, about 10 per cent of the baryonic mass, is traced by
\ion{O}{vi} absorption lines with equivalent width $W>20$ m\AA. They expect
about five \ion{O}{vi} absorption lines per unit redshift with $W>35$
m\AA\ and about 0.5 per unit redshift with $W>350$ m\AA.
\citet{Fang2001} estimate that most \ion{O}{vi} absorption lines are
correlated to filamentary regions with overdensities of 5--100 and
temperatures of a few times $10^5$ K.

The equivalent width of absorption systems in the ultraviolet region
can be useful to distinguish a prevalently photoionised from a
prevalently collisionally ionised IGM.  Collisional ionisation
dominates in the denser and warmer regions of the IGM and is
responsible for most lines with $W\ga 40$ m\AA. On the other hand,
photoionisation is predominant in the cooler and less dense regions
and usually produces narrow absorption lines with $W\la 40$
m\AA. Essentially all lines with equivalent width $W>80$ m\AA\ are due
to collisionally ionised gas \citep{Fang2001}.

\citet{richter2006} investigate the origin of the broad
\ion{H}{i} \lya\ absorbers (BLA) with Doppler parameters $b>40$ km
s\uno\ seen in UV spectra from the \hst\ and
\fuse\ (\citealt{richter2004}; \citealt{richter2008} - Chapter 4, this volume). They find that such absorbers are associated with WHIM gas
with temperatures in the range $2.5 \times 10^4 < T < 1.6 \times 10^6$
K, that represents about a quarter of the gas mass in the simulation
at $z=0$. Less than one third of the broad \ion{H}{i} absorbers are
associated with cooler gas with $T < 2 \times 10^4$ K, whose
broadening is produced by non-thermal processes such as turbulence and
velocity structures in the IGM.

\section{Sunyaev-Zel'dovich signal of the WHIM}
\label{sz}

It has been suggested that ionised WHIM gas could possibly be detected
through the thermal Sunyaev-Zel'dovich effect it produces in the CMB
(e.g. \citealt{croft2006}; \citealt{cao2006}; \citealt{Hallman2007};
\citealt{bregman2007}).  When the CMB radiation crosses a hot ionised
medium such as the WHIM, the energy of the passing photons can be
increased by collisions with the free electrons in the WHIM. This
produces a local increase of the CMB temperature, with an increase
proportional to the line integral of the pressure of the hot gas.
Although the signal arising from the hot gas in large clusters has
been clearly identified, a weaker signal associated with WHIM gas has
not yet been reliably detected in current CMB data
(\citealt{carlos2004}; \citealt{hansen2007}).

\citet{Hallman2007} estimate that the WHIM could contribute on average
between 4 and 12 per cent of the integrated Sunyaev-Zel'dovich effect
(SZE) signal $Y$ for individual sources identified in upcoming
surveys.  The WHIM is a particularly relevant source of contamination for
observations of clusters of galaxies, because it would strongly affect
the calibration of the cluster $Y-M$ relationship. Unless the WHIM signal
can be unambiguously modelled and separated from the cluster signal,
it introduces a bias and a source of scatter difficult to quantify in
the $Y-M$ relationship. Ultimately, this would limit the reliability
of the $Y-M$ relationship for estimating cosmological parameters.

\section{Intergalactic magnetic fields}
\label{bfield}

It has been suggested that cosmic shear and gravitational shocks might
be able to amplify the magnetic fields that permeate the large-scale
structure (e.g. \citealt{dolag1999}; \citealt{min01a};
\citealt{marcus2005}). Given the strong interconnection between the
origin of the WHIM and gravitational shocks, it is possible for the
WHIM to be magnetised, although it might be difficult to predict to
what level.

While magnetic fields have been unambiguously detected in the cores of
galaxy clusters, few observations have attempted to establish the
presence of magnetised plasma in low density regions such as groups
and filaments. However, there is now compelling evidence that magnetic
fields might exist beyond clusters, and in particular in the WHIM.
\citet{kim} claim the detection of an extended magnetic field in the
region of the Coma supercluster. They observe radio-synchrotron
radiation from a region with an enhanced number density of galaxies,
seeming to indicate the presence of a group of galaxies, perhaps in
the process of merging with the Coma cluster.  Evidence for
intergalactic magnetic fields in the region ZwCL~2341+0000 at $z \sim
0.3$, where galaxies seems to lie along a filament, has been reported
by \citet{bagchi}.  In both cases, the inferred strength of the
magnetic field is of order $0.01 - 0.1$ $\mu$G.

Several mechanisms have been proposed to explain the injection of seed
magnetic fields into the intracluster and intergalactic media, later
to be amplified by cosmic shear and gravitational shocks. Among them,
primordial magnetic fields, jets and radio lobes emerging from radio
galaxies (e.g. \citealt{hoyle}; \citealt{chak}; \citealt{ensslin1997})
and galactic winds (e.g. \citealt{kron1999}; \citealt{bertone}).
Simulations (\citealt{kron1999}; \citealt{kron2001};
\citealt{bertone}) show that magnetic fields can be efficiently
transported into the IGM and fill most of the universe by
$z=0$. \citet{bertone} predict that galactic winds alone could be
responsible for IGM magnetic fields of order $10^{-6} - 10^{-2}$
$\mu$G, in agreement with observations.

\section{Conclusions}
\label{conclusion}

The search for the ``missing baryons'' has opened the way to our
understanding of the most elusive baryonic component in the universe:
the hot diffuse gas that traces the cosmic web.

Given the paucity of observational findings, most of what we know
about the WHIM is based on numerical simulations.  These
predict that at $z<1$ up to 50 per cent of the IGM has been heated to
temperatures in the range $T \sim 10^5 - 10^7$ K. While most of the
heating is provided by gravitational accretion shocks propagating out
of regions undergoing 
gravitational collapse, non-gravitational processes such as galactic
winds and AGN feedback may play an important role in high density
regions, mostly by preventing cooling and regulating the intensity of the
X-ray background.

The observability of the WHIM by future facilities has received a lot
of attention and a wealth of predictions have been made based on
numerical results.  The detection of the WHIM seems to be extremely
challenging, if not impossible, with current instruments and a new
generation of X-ray telescopes with higher sensitivity and spectral
resolution, such as \constellation\ and \xeus, is needed to study the
WHIM in absorption. In addition
to high sensitivity and spectral resolution, high spatial resolution
on a large field of view, as proposed for the Explorer of Diffuse
emission and Gamma-ray burst Explosions (\edge), would be ideal for
mapping the WHIM emission on the sky.

Although most theoretical predictions claim optimistically that we
will be able to detect the WHIM in absorption with \constellation\ and
\xeus, and maybe in emission with some luck, there is a long way to go
before observations can confirm the simulation predictions, or show us
a different picture altogether of what and where the WHIM is.

\begin{acknowledgements}

We are grateful to Frits Paerels and Philipp Richter for a careful
reading of the manuscript. 
We thank ISSI (Bern) for support of the team ``Non-virialized
X-ray components in clusters of galaxies''. This work was supported by a
Marie Curie Excellence Grant MEXT-CT-2004-014112. SB acknowledges support from STFC and from a NSF grant AST-0507117. 

\end{acknowledgements}

\bibliographystyle{aa}

\end{document}